\documentclass[twocolumn]{aastex63}

\usepackage[normalem]{ulem}
\usepackage{hyperref}
\usepackage{bm,bbm,amssymb,amsmath}
\usepackage{graphicx}
\usepackage{comment}
\usepackage{natbib}

\usepackage[inline,nolabel,final]{showlabels} 
\newcommand{\figlabel}[1]{\label{#1}} 

\newcommand{\Msun}{\ensuremath{\,M_{\odot}}}

\begin{document}

\title{The Mass Distribution of Neutron Stars in Gravitational-wave Binaries}

\author{Philippe Landry}
\author{Jocelyn S. Read}
\affiliation{Nicholas \& Lee Begovich Center for Gravitational-Wave Physics \& Astronomy, California State University, Fullerton, \\ 800 N State College Boulevard, Fullerton, CA 92831, USA}

\begin{abstract}
The discovery of two neutron star--black hole coalescences by LIGO and Virgo brings the total number of likely neutron stars observed in gravitational waves to six. We perform the first inference of the mass distribution of this extragalactic population of neutron stars. In contrast to the bimodal Galactic population detected primarily as radio pulsars, the masses of neutron stars in gravitational-wave binaries are thus far consistent with a uniform distribution, with a greater prevalence of high-mass neutron stars. The maximum mass in the gravitational-wave population agrees with that inferred from the neutron stars in our Galaxy and with expectations from dense matter. \\
\end{abstract}

\section{Introduction}

 To date, the vast majority of neutron star (NS) masses have been measured through the binary interactions of Galactic radio pulsars \citep{OzelFreire2016}. Beginning with \citet{ThorsettChakrabarty1999}, population-level analyses have been performed on these Galactic pulsars, revealing a mass distribution with a double-peaked structure~\citep{2016arXiv160501665A, AlsingSilva2018,FarrChatziioannou2020,ShaoTang2020}. The double NSs among these Galactic observations, however, have only been found with both component masses near 1.35\Msun~\citep{OzelPsaltis2012,KiziltanKottas2013}. 
 
The first NS binary observed in gravitational waves (GWs), GW170817 \citep{GW170817}, fell neatly into the mass range anticipated from these radio observations. On the other hand, GW190425~\citep{GW190425}, the second low-mass GW binary discovered, illustrated that the compact object population detected with GWs need not be the same as that observed in our Galaxy: its total mass is a 5$\sigma$ outlier relative to known Galactic double NSs. With the recent neutron star--black hole (NSBH) discoveries GW200105 and GW200115~\citep{NSBHs},  LIGO~\citep{aLIGO} and Virgo~\citep{aVirgo} have now recorded four confirmed signals whose progenitor likely contains at least one NS. 
In this Letter, we consider these to be six extragalactic NSs known through GW astronomy, and investigate the collective properties of this incipient NS population.  We follow the hierarchical Bayesian population inference framework of \citet{Mandel2010} and \citet{MandelFarr2019}, which accounts for statistical uncertainty in the measured parameters of each GW signal and for selection effects in the observed population. This kind of population inference has a history of applications inside \citep[e.g.][]{O1O2pop,RouletVenumadhav2020,O3apop} and outside \citep[e.g.][]{Loredo2004,HoggMyers2010,ForemanMackeyHogg2014} of GW astronomy.

One of the chief aims of this work is to compare the GW and Galactic NS mass distributions. We also investigate the maximum mass inferred for the extragalactic NS population, the NS pairing probability for assembling BNSs, and the effect of including or excluding an outlier observation, GW190814 \citep{GW190814}; the nature of its 2.6\Msun~secondary component is challenging to interpret, as it falls into the mass gap between known high-mass NSs \citep{2020NatAs...4...72C} and low-mass black holes \citep[BHs;][]{2019Sci...366..637T}.

On the whole, we find that a minimalistic model consisting of a uniform mass distribution with random pairing into BNSs is adequate for explaining the population of NSs observed so far with GWs. Moreover, these first observations already reveal interesting differences between this extragalactic NS population and that which we observe in our own Galaxy. \\

\section{Neutron Stars in Gravitational-wave Binaries}

The set of observations informing our population inference consists of the BNS mergers GW170817 and GW190425, and the NSBH mergers GW200105 and GW200115. We separately consider treating GW190814 as another NSBH merger, given the uncertainty in the nature of its secondary component. GWs from compact binary inspirals such as these provide accurate measurements of their source's primary and secondary masses, $m_{1,2}$, as well as its luminosity distance $D_L$. These measurements, in the form of a three-dimensional marginal likelihood $P(d|m_1,m_2,D_L)$, are the observational input for our population inference. The marginal likelihood can be obtained from the full GW likelihood $P(d|m_1,m_2,D_L,\boldsymbol{\theta})$ by marginalizing over nuisance parameters $\boldsymbol{\theta}$, such as the source's inclination and spins. Although our NS population model is independent of $D_L$, we must retain the luminosity distance dependence in the likelihood in order to model GW selection effects.

For each of the observations, we calculate the marginal likelihood from the GW posterior samples released by the LIGO--Virgo Collaboration.\footnote{We use the minimal spin-assumption (``high spin'') posterior samples from \url{https://dcc.ligo.org/LIGO-P1800061/public} for GW170817, \url{https://dcc.ligo.org/LIGO-P2000026/public} for GW190425, \url{https://dcc.ligo.org/LIGO-P2100143/public} for GW200105 and GW200115, and \url{https://dcc.ligo.org/LIGO-P2000183/public} for GW190814. These samples are available through the Gravitational-Wave Open Science Center~\citep[\url{https://www.gw-openscience.org};][]{VallisneriKanner2015,2021SoftX..1300658A}.} That is, we weight every posterior sample $(m_1,m_2,D_L)$ by a factor of $1/\pi_\emptyset(m_1,m_2,D_L)$ to convert from the prior $\pi_\emptyset(m_1,m_2,D_L)$ adopted for GW parameter estimation, which is uniform in redshifted masses $m_{1,2}^z = (1+z) m_{1,2}$ and quadratic in the luminosity distance, to the uniform-in-$m_{1,2}$-$D_L$ prior required for the likelihood. Gaussian kernel density estimates of the one-dimensional marginal component mass likelihoods constructed from these weighted samples are plotted in the upper panel of Fig.~\ref{fig:flatmass} to illustrate the mass measurements that drive the population inference.

\section{Modeling the Neutron Star Mass Distribution} \label{sec:models}

Motivated by a few basic considerations, we investigate several simple models for the NS mass distribution. Firstly, causality arguments place an upper bound of $\sim 3 \Msun$ on the maximum NS mass~\citep{RhoadesRuffini1974,KalogeraBaym1996}. Plausible supernova formation channels do not produce NSs below 1\Msun~\citep{2012ApJ...749...91F,2020ApJ...896...56W} and the standard LIGO/Virgo searches target masses greater than 1\Msun~\citep{2016ApJ...832L..21A,2016CQGra..33u5004U,2021PhRvD.103h4047M}. 
Our population models are consequently formulated for $m \in [1,3)\Msun$.

Second, although there is evidence that recycled and nonrecycled pulsars in the Galactic population originate from different mass distributions~\citep{OzelPsaltis2012,FarrowZhu2019}, we do not expect the current limited set of GW mass measurements to have the resolving power to discriminate between different subpopulations. We therefore assume that all NSs in BNSs and NSBHs originate from a common mass distribution $\pi(m|\boldsymbol{\lambda})$.

Third, again because of the relatively small number of NS observations at hand, we prefer simple shapes---uniform or Gaussian---for the common mass distribution. We also consider a sum of two Gaussians to facilitate a direct comparison with the bimodal Galactic population \citep{2016arXiv160501665A,AlsingSilva2018,FarrChatziioannou2020,ShaoTang2020}.

Given these considerations, we adopt one of three models for the basic NS mass distribution:

\begin{widetext}
\begin{subequations} \label{ns_mass_models}
\begin{align}
    \pi_\textsc{u}(m|m_{\rm min},m_{\rm max}) :=& U(m|m_{\rm min},m_{\rm max}) , \label{flat} \\
    \pi_\textsc{n}(m|\mu,\sigma,m_{\rm min},m_{\rm max}) :=& \mathcal{N}(m|\mu,\sigma)U(m|m_{\rm min},m_{\rm max})/A , \label{peak} \\
    \pi_\textsc{nn}(m|\mu,\sigma,\mu',\sigma',w,m_{\rm min},m_{\rm max}) :=& \left[w \mathcal{N}(m|\mu,\sigma)/B + (1-w) \mathcal{N}(m|\mu',\sigma')/C\right] U(m|m_{\rm min},m_{\rm max}) , \label{bimodal}
\end{align}
\end{subequations}
\end{widetext}
and subject the population parameters $\boldsymbol{\lambda}$ to the constraint $1 \Msun \leq m_{\rm min} < \mu < \mu' < m_{\rm max} < 3 \Msun$. The normalization constants $A$, $B$, and $C$ in Eqs.~\eqref{peak}-\eqref{bimodal} ensure that $\int \pi(m|\boldsymbol{\lambda}) \, dm = 1$.

\begin{figure}[htb!]
\centering
\includegraphics[width=0.98\columnwidth,trim={0 0 0 0},clip]{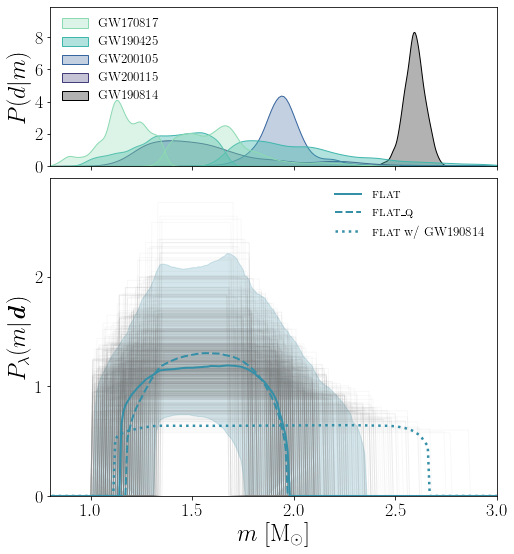}
\caption{Measured masses and inferred mass distribution for NSs in GW binaries. Top: marginal one-dimensional mass likelihoods $P(d|m)$ for the NSs in the BNS mergers GW170817 and GW190425,  the NSBH mergers GW200105 and GW200115, and the candidate NSBH merger GW190814. Bottom: median and symmetric 90\% confidence interval of the mass distribution inferred for the \textsc{flat} model. The gray traces plot 1000 independent samples from the \textsc{flat} population posterior. The dashed line is the median for the \textsc{flat\_q} model. The dotted line is the \textsc{flat} median when GW190814 is interpreted as an NSBH.
\figlabel{fig:flatmass}}
\end{figure}

The astrophysical processes of stellar evolution and binary interaction that eventually result in a merging system containing one or two NSs have significant complexity and uncertainty \citep{2012ApJ...759...52D,2019ApJ...883L..45F, 2021arXiv210302608B, 2021MNRAS.500.1380M, 2021MNRAS.502.4877S}. In the absence of a definite, astrophysically motivated prescription for the pairing function that connects the masses of the two components in a GW binary, we suppose that the pairing probability is a power law in the mass ratio $q = m_2/m_1$. This phenomenological model is a good descriptor of pairing in the binary BH population~\citep{KovetzCholis2017,O1O2pop,FishbachHolz2020,O3apop}. The binary mass distribution characteristic of BNSs and NSBHs is thus taken to be

\begin{equation} \label{bns_mdistr}
\pi(m_1,m_2|\boldsymbol{\lambda}) \propto \begin{cases}
           \pi(m_1|\boldsymbol{\lambda}) \pi(m_2|\boldsymbol{\lambda}) q^\beta \; &\text{if BNS} \\
          \pi_{\textsc{bh}}(m_1) \pi(m_2|\boldsymbol{\lambda}) q^{\beta_{\textsc{bh}}} \; &\text{if NSBH} , \\
     \end{cases}
\end{equation}
with separate pairing parameters $\beta$ and $\beta_{\textsc{bh}}$ for the BNSs and NSBHs. We assume definite \textit{a priori} classifications for the GW sources.\footnote{If we were instead to classify the sources probabilistically based on their measured parameters, Eq.~\eqref{bns_mdistr} would involve a mixing fraction between BNSs and NSBHs that depends on their relative detection rate.} Because there are so few observations of BHs in NSBH systems, we do not attempt to infer the BH mass distribution $\pi_\textsc{bh}(m)$; rather, we fix it to be uniform and nonoverlapping with the NS mass distribution: $\pi_{\textsc{bh}}(m) = U(m|3\Msun, 30\Msun)$. In practice, the criterion in Eq.~\eqref{bns_mdistr} that distinguishes BNSs from NSBHs is $m_1 < 3\Msun$. The BH mass distribution's upper bound is chosen to encompass the measurement for GW190814's primary component. We also choose to fix the NSBH pairing parameter to $\beta_\textsc{bh} = 0$, which corresponds to a minimal assumption of random pairing. These choices do not strongly impact the inferred NS mass distribution, which is driven by the NS mass model and measurements. Indeed, we have tested a different BH mass model---a power law $\propto m^{-2}$---and a different NSBH pairing parameter---$\beta_\textsc{bh} = 2$, a value that lies within the 90\% credible region preferred by the entire compact binary population in the second LIGO--Virgo GW transient catalog \citep{GWTC2,O3apop}---and found a negligible impact on the inferred NS mass distribution.

We also take Eq.~\eqref{bns_mdistr} to be independent of the GW binary's luminosity distance. This amounts to an assumption that the NS and BH populations do not evolve with distance, which is a reasonable approximation for nearby signals~\citep{2018ApJ...863L..41F, 2019MNRAS.487....2M}.

Hence, the binary mass model~\eqref{bns_mdistr} is specified by a choice of basic NS mass distribution $\pi(m|\boldsymbol{\lambda})$ and a choice of BNS pairing parameter $\beta$. In light of limited existing GW observations, we make a minimalistic assumption for the latter: we assume random pairing, i.e.~$\beta=0$. (We will revisit this assumption in Section~\ref{sec:results}.) Our three fiducial models, which we call \textsc{flat}, \textsc{peak} and \textsc{bimodal} in Table~\ref{tab:metrics}, therefore differ only by the basic NS mass distribution shape they adopt: respectively uniform, Gaussian, and double Gaussian. 

\begin{table*}[htb!]
    \centering
    \begin{tabular}{lcccccccccc} \hline\hline
        Model & $\pi(m|\boldsymbol{\lambda})$ & $\beta$ & $m_{\rm min}\,(M_{\odot})$ & $m_{\rm max}\,(M_{\odot})$ & $\mu\,(M_{\odot})$ & $\sigma\,(M_{\odot})$ & $\mu'\,(M_{\odot})$ & $\sigma'\,(M_{\odot})$ & $w$ & $\Delta {\rm AIC}$ \\ \hline
        \textsc{flat} & \textsc{u} & 0 & $1.1^{+0.2}_{-0.1}$ & $2.0^{+0.4}_{-0.3}$ & - & - & - & - & - & 0.0 \\ 
        \textsc{peak} & \textsc{n} & 0 & $1.1^{+0.2}_{-0.1}$ & $2.1^{+0.7}_{-0.3}$ & $1.5^{+0.5}_{-0.3}$ & $1.0^{+0.9}_{-0.7}$ & - & - & - & 3.7 \\ 
        \textsc{bimodal} & \textsc{nn} & 0 & $1.0$ & $2.1^{+0.7}_{-0.3}$ & $1.3^{+0.5}_{-0.3}$ & $1.0^{+0.9}_{-0.8}$ & $1.7^{+0.6}_{-0.4}$ & $1.0^{+0.9}_{-0.9}$ & $0.5^{+0.5}_{-0.4}$ & 5.5 \\ 
        \textsc{flat\_q} & \textsc{u} & 2 & $1.2^{+0.1}_{-0.2}$ & $2.0^{+0.3}_{-0.3}$ & - & - & - & - & - & 1.1 \\ 
        \textsc{peak\_q} & \textsc{n} & 2 & $1.1^{+0.2}_{-0.1}$ & $2.0^{+0.7}_{-0.2}$ & $1.5^{+0.5}_{-0.3}$ & $1.0^{+0.9}_{-0.7}$ & - & - & - & 4.8 \\ 
        \textsc{bimodal\_q} & \textsc{nn} & 2 & $1.0$ & $2.1^{+0.7}_{-0.3}$ & $1.3^{+0.5}_{-0.3}$ & $0.9^{+1.0}_{-0.8}$ & $1.7^{+0.5}_{-0.4}$ & $1.0^{+0.9}_{-0.9}$ & $0.5^{+0.5}_{-0.4}$ & 8.1 \\ 
        \hline\hline
    \end{tabular}
    \caption{Inferred Population Parameters for the NS Mass Distribution According to Each of the Models Considered. Medians and symmetric 90\% credible intervals of the marginal one-dimensional posterior distributions are reported. The minimum mass is fixed to 1 \Msun~for the \textsc{bimodal} and \textsc{bimodal\_q} models.
    The AIC statistic, used for ranking the models, is also reported relative to the \textsc{flat} model.
    \figlabel{tab:metrics}}
\end{table*}

\section{Gravitational-wave Population Inference}
Bayesian population inference supplies a prescription for evaluating the likelihood of our set $\boldsymbol{d}$ of GW observations, given a population model $\pi(m_1,m_2|\boldsymbol{\lambda})$ that depends on  unknown population parameters $\boldsymbol{\lambda}$ \citep{Mandel2010,MandelFarr2019}. Formally,

\begin{align} \label{pop_like}
    P(\boldsymbol{d}|\boldsymbol{\lambda}) = \prod_i \frac{1}{\zeta(\boldsymbol{\lambda})} \int \big[& P(d_i|m_1,m_2,D_L)\pi(m_1,m_2|\boldsymbol{\lambda}) \nonumber \\
    &\times P(D_L)\big] \, dm_1 \, dm_2 \, d D_L ,
\end{align}
where the three-dimensional marginal likelihood $P(d|m_1,m_2,D_L)$ is the observational input from the GWs. The detection fraction

\begin{align} \label{sensitivity}
    \zeta(\boldsymbol{\lambda}) = \int \big[ &P_{\rm det}(m_1,m_2,D_L)\pi(m_1,m_2|\boldsymbol{\lambda}) \nonumber \\
    & \times P(D_L) \big] \, dm_1 \, dm_2 \, dD_L,
\end{align}
which acts as a normalization in Eq.~\eqref{pop_like}, captures GW selection effects. The selection function $P_{\rm det}(m_1,m_2,D_L)$ gives the probability of detection for a GW signal as a function of its source properties. We adopt a simple analytic model of a hard signal-to-noise (SNR) ratio threshold following \citet{ChatziioannouFarr2020}, which produces a selection function that scales like ${\mathcal{M}_z}^{5/2}$, where $\mathcal{M}_z = (1+z) (m_1 m_2)^{3/5}/(m_1 + m_2)^{1/5}$ is the redshifted chirp mass.  
In Eqs.~\eqref{pop_like}-\eqref{sensitivity}, the distance prior $P(D_L) \propto {D_L}^2$ encodes the quadratic increase in a GW survey's ratio of sensitive volume to distance in the local universe.

The posterior probability of a set of population parameters $\boldsymbol{\lambda}$ within a given population model follows from Eq.~\eqref{pop_like} via Bayes' theorem: $P(\boldsymbol{\lambda}|\boldsymbol{d}) \propto P(\boldsymbol{d}|\boldsymbol{\lambda}) P(\boldsymbol{\lambda})$, where $P(\boldsymbol{\lambda})$ is their prior distribution. The model's prediction for the intrinsic NS mass distribution is obtained by marginalizing its binary mass distribution over the posterior uncertainty in $\boldsymbol{\lambda}$ and averaging over the individual $m_1$ and $m_2$ distributions:

\begin{align}
    P_\lambda(m|\boldsymbol{d}) =& \frac{1}{2} \int \pi(m_1,m|\boldsymbol{\lambda}) P(\boldsymbol{\lambda}|\boldsymbol{d}) \,d\boldsymbol{\lambda} \, dm_1 \nonumber \\ &+ \frac{1}{2} \int \pi(m,m_2|\boldsymbol{\lambda}) P(\boldsymbol{\lambda}|\boldsymbol{d}) \,d\boldsymbol{\lambda}\,dm_2 . \label{ppopd}
\end{align}
The corresponding prediction for the \emph{observed} NS mass distribution $P_\lambda^{\rm obs}(m|\boldsymbol{d})$---i.e., the one filtered through selection effects---can be obtained by inserting factors of $\zeta(\boldsymbol{\lambda})$ within the integrals. \\

\section{Results and Discussion} \label{sec:results}
For each of the models described in Sec.~\ref{sec:models}, we follow the previous section's formalism and evaluate $P(\boldsymbol{\lambda}|\boldsymbol{d})$ with a Markov Chain Monte Carlo algorithm implemented with \texttt{emcee}~\citep{ForemanMackeyHogg2013}. We use the marginal likelihoods from GW170817, GW190425, GW200105, and GW200115 as the observational input. We sample from uniform priors over the population parameters, subject to the constraints listed after Eq.~\eqref{ns_mass_models}, within the ranges $m_{\rm min} \in [1.0,1.5)$\Msun, $m_{\rm max} \in [1.5,3.0)$\Msun, $\mu, \mu' \in [1.0,3.0)$\Msun, $\sigma, \sigma' \in [0.01,2.00]$\Msun~and $w \in [0.0,1.0]$. For the \textsc{bimodal} model, we fix $m_{\rm min} = 1.0$ to reduce the dimensionality of the population parameter space. The detection fraction $\zeta(\boldsymbol{\lambda})$ for each realization of the population model is evaluated via Monte Carlo integration of Eq.~\eqref{sensitivity}, and likewise the inferred mass distribution~\eqref{ppopd} is computed from the posterior distribution $P(\boldsymbol{\lambda}|\boldsymbol{d})$.

\paragraph{Inferred neutron star mass distribution}

We first examine the inferred mass distribution for the three fiducial models described above to discern the key characteristics of the NS population in GW binaries. The median and symmetric 90\% confidence interval of the mass distribution inferred for the \textsc{flat}, \textsc{bimodal} and \textsc{peak} models are plotted in Figs.~\ref{fig:flatmass}, \ref{fig:galactic} and \ref{fig:peak}, respectively. These plots also feature traces of individual realizations of the mass distribution drawn from the population posterior. Table~\ref{tab:metrics} reports the median and symmetric 90\% confidence interval of the one-dimensional marginal posterior distribution for each population parameter.

\begin{figure}[tb!]
\centering
\includegraphics[width=0.98\columnwidth,trim={0 0 0 0},clip]{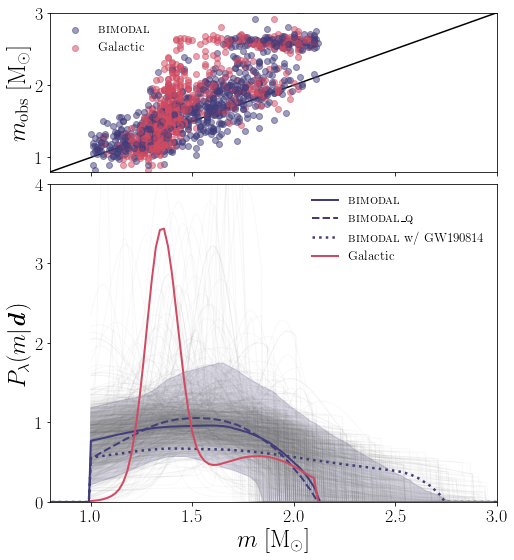}
\caption{Comparison between the inferred mass distributions for NSs in GW binaries and NSs in the Galaxy. Top: posterior predictive check of the \textsc{bimodal} and Galactic population models. One hundred realizations of the observed NS masses are plotted against 100 samples from the predicted \textsc{bimodal} and Galactic mass distributions $P^{\rm obs}_\lambda(m|\boldsymbol{d})$. The black line with unit slope indicates perfect agreement between the model and the observations; a model that overpredicts (underpredicts) the number of NS observations of a given mass will systematically produce scatter points below (above) the line. Bottom: same as Fig.~\ref{fig:flatmass}, but for the \textsc{bimodal} model as compared to the median of the \citet{FarrChatziioannou2020} Galactic mass distribution.
\figlabel{fig:galactic}}
\end{figure}

\begin{figure}[htb!]
\centering
\includegraphics[width=0.98\columnwidth,trim={0 0 0 0},clip]{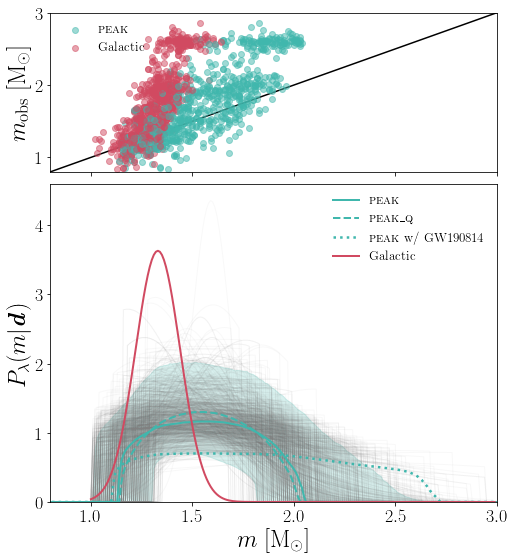}
\caption{Same as Fig.~\ref{fig:galactic}, but for the \textsc{peak} model and the best-fit Gaussian Galactic double NS mass distribution from \citet{KiziltanKottas2013}.
\figlabel{fig:peak}}
\end{figure}

The inferred \textsc{flat} mass distribution extends from approximately 1--2\Msun, with considerable uncertainty in the minimum and maximum mass cutoffs, as expected from only a handful of GW observations; \citet{ChatziioannouFarr2020} estimate that 50 BNS signals are needed to determine both cutoffs to within 0.2\Msun. Nonetheless, the observations constrain the $m_{\rm min}$ and $m_{\rm max}$ parameters away from the prior, disfavoring in particular the largest maximum mass values allowed \textit{a priori}. The lower bound on $m_{\rm max}$ is driven by the precise mass measurement for GW200105's 1.9\Msun~secondary, which must be accommodated by the mass distribution; this causes the pileup in traces that corresponds to the precipitous fall in the median at 2\Msun~in Fig.~\ref{fig:flatmass}. The overlap of the less sharply peaked likelihoods in the 1.1--1.6\Msun~range is responsible for the slightly more gradual initial rise in the median. 

The mass distributions inferred with the \textsc{bimodal} and \textsc{peak} models are very similar. Both are unimodal on average, with a broad peak at $\sim 1.6\Msun$; although many \textsc{bimodal} traces have two peaks, the peak locations are relatively unconstrained by the observational data, which washes out the bimodality in the median. Similarly, despite many individual realizations of the mass distribution having a sharp high-mass cutoff, the GW observations primarily bound $m_{\rm max} \gtrsim 2\Msun$ for these models, allowing the median to taper off smoothly at high masses. At the low-mass end, we observe a similar tapering in the \textsc{peak} mass distribution, while the \textsc{bimodal} mass distribution is dominated by the imposed sharp cutoff at $m_{\rm min} = 1\Msun$.

Besides examining the NS mass distribution predicted by each of the three models, we can attempt to determine which model is preferred by the observations. We adopt the Akaike information criterion~\citep[AIC;][]{AKAIKE19813}, ${\rm AIC} = 2n - 2\log{P(\boldsymbol{d}|\boldsymbol{\hat{\lambda}})}$, as the metric for this comparison. The AIC is simply the (logarithm of the) maximal likelihood $P(\boldsymbol{d}|\boldsymbol{\hat{\lambda}})$ for a given population model, where $\boldsymbol{\hat{\lambda}}$ are the maximum-likelihood parameters, plus a correction term that penalizes the model according to its number of free parameters $n$. The smaller the AIC, the greater the data's preference for the model. The AIC is used for model selection in the astrophysics literature \citep[e.g.][]{ShiHuang2012,KrishakDesai2020}, with $\exp{(\Delta {\rm AIC}/2)}$ corresponding to the relative likelihood of model $B$ compared to model $A$ for $\Delta {\rm AIC} = {\rm AIC_A}-{\rm AIC_B}$. Values of $\Delta {\rm AIC} > 6$ are considered significant for model selection~\citep{Liddle2009}.

To rank our population models, for each one we identify the mass distribution realization with the largest likelihood among our discrete population posterior and calculate $\Delta{\rm AIC}$ relative to the \textsc{flat} model. Based on the results listed in Table~\ref{tab:metrics}, we find that the \textsc{peak} and \textsc{bimodal} models are moderately disfavored by the data, in part from their larger number of free parameters ($n = 4$ and $n = 6$, respectively). The more parsimonious ($n = 2$) \textsc{flat} model is preferred. 

Thus, a uniform NS mass distribution is sufficient to fit current GW observations. There is no clear evidence of bimodality in this GW population of NSs, nor is there unambiguous evidence of a sharp minimum or maximum mass cutoff; these latter features only appear in the inferred \textsc{flat} mass distribution, which includes them by construction.

\paragraph{Comparison with the Galactic neutron star population}

The Galactic population of NSs, observed via electromagnetic astronomy, has been studied extensively \citep{OzelFreire2016}. The double NS subpopulation appears to conform to a sharply peaked Gaussian mass distribution with $\mu = 1.33\Msun$ and $\sigma \approx 0.11\Msun$~\citep{KiziltanKottas2013}, at least when assuming a common distribution for recycled and nonrecycled pulsars~\citep{OzelPsaltis2012,FarrowZhu2019}. The full NS population is characterized by a bimodal mass distribution~\citep{2016arXiv160501665A,AlsingSilva2018,FarrChatziioannou2020,ShaoTang2020}. A recent study \citep{FarrChatziioannou2020} found Gaussian peaks at $\mu = 1.35^{+0.04}_{-0.03}\Msun$ and $\mu' = 1.8^{+0.6}_{-0.2}\Msun$, with respective widths $\sigma = 0.08^{+0.03}_{-0.03}\Msun$ and $\sigma' = 0.3^{+0.3}_{-0.1}\Msun$, a maximum mass cutoff at $m_{\rm max} = 2.3^{+0.8}_{-0.3}\Msun$, and $w = 0.7^{+0.1}_{-0.2}$.\footnote{We use the population posterior samples from \citet{FarrChatziioannou2020} made available at \url{https://github.com/farr/AlsingNSMassReplication}.} Comparing these two Galactic models with our inferred \textsc{peak} and \textsc{bimodal} mass distributions, which are directly analogous, allows us to investigate the differences between the Galactic and GW populations of NSs.

In Fig.~\ref{fig:galactic}, we compare the double-Gaussian mass distribution from \citet{FarrChatziioannou2020} with the \textsc{bimodal} model. Unlike the Galactic mass distribution, the one inferred from GWs is unimodal, predicting far fewer low-mass and moderately more high-mass NSs in the population. The \textsc{bimodal} mass distribution may also support a gentler maximum mass cutoff. The apparent difference at the low-mass end of the Galactic and GW mass distributions is an artifact of our choice to fix $m_{\rm min} = 1\Msun$ in the \textsc{bimodal} model---consistent with our prior bounds---whereas \citet{FarrChatziioannou2020} set $m_{\rm min} = 0$.

Similarly, in Fig.~\ref{fig:peak}, we compare the Gaussian double NS mass distribution from \citet{KiziltanKottas2013} against the \textsc{peak} model. The peak in the latter mass distribution is much broader than the Galactic one and occurs at a higher mass scale. Hence, the \textsc{peak} model predicts many more heavy double NSs in the population than are observed in our Galaxy. This is consistent with the discovery of GW190425, whose total mass of approximately $3.4\Msun$ is already recognized as an outlier from the observed Galactic double NS population~\citep{GW190425}.

To reinforce the conclusion that the observed populations of NSs in the Galaxy and in GW binaries are distinct, we perform a posterior predictive check of the models. Essentially, we test whether the Galactic models provide an equally good fit to the observed GW population of NSs as our \textsc{bimodal} and \textsc{peak} models. To do so, we draw one mass sample from each of the GW marginal mass likelihoods to constitute a realization of the observed NS population, and we draw the same number of samples from either the GW or Galactic mass distribution $P_\lambda^{\rm obs}(m|\boldsymbol{d})$. This constitutes the prediction for the detected NS population, accounting for selection effects. Iterating over 100 instances of this procedure, we pair up the observed and predicted masses in sequence of increasing mass, and plot the resulting scatter points in the upper panels of Figs.~\ref{fig:galactic} and \ref{fig:peak}. A model that fits the observations perfectly would have its scatter points distributed along the line of unit slope. Despite the considerable statistical uncertainty in the mass likelihoods and inferred mass distribution for the GW population models, we see that their scatter points are generally oriented along this line. On the other hand, the scatter points for both Galactic models skew above the line, indicating that they underpredict the number of high-mass NSs observed with GWs. This is especially apparent for the double NS model from \citet{KiziltanKottas2013} in Fig.~\ref{fig:peak}.

Thus, a mass model distinct from the existing Galactic ones is indeed required to accurately describe the detected GW population of NSs. We account for the GW selection bias toward detecting heavy masses, so this may reflect a difference in the observed properties of Galactic pulsars vs.~the intrinsic properties of merging compact binaries as a result of different binary evolution pathways. For example, \citet{GalaudageAdamcewicz2021} suggest that the rapid merger of high-mass systems reduces their radio visibility, which could explain the relative lack of such systems seen in our Galaxy. More generally, the properties of the observed Galactic population are affected by radio selection effects \citep{2020MNRAS.494.1587C}, which we do not explore in this work. It may be possible to reconcile the two observed populations by building additional parameters  such as the spins or the orbital period into the model \citep{Kruckow2020}.

\paragraph{Maximum mass}

The best-constrained population parameter in each model is the maximum mass, $m_{\rm max}$. We investigate whether the maximum mass in the GW population of NSs agrees with (1) the maximum mass in the Galactic population, and (2) the maximum Tolman--Oppenheimer--Volkoff (TOV) mass supported by the NS equation of state (EOS). A discrepancy with the former could be informative about differences between GW and electromagnetic selection effects; a discrepancy with the latter could indicate that the NS mass spectrum is limited by the astrophysical formation channel rather than the EOS.

In Fig.~\ref{fig:mmax}, we plot the $m_{\rm max}$ posterior distribution inferred with the \textsc{flat} and \textsc{bimodal} models. Both posteriors peak around $2\Msun$, but the \textsc{bimodal} model has support for $m_{\rm max}$ as large as $3\Msun$ because its allows for tapering instead of a sharp cutoff. The \textsc{flat} model predicts a maximum mass of $2.0^{+0.4}_{-0.3}\Msun$, while the \textsc{bimodal} model predicts $2.1^{+0.7}_{-0.3}\Msun$. The lower bound on $m_{\rm max}$ is driven by the well-resolved mass of the secondary in GW200105 in both cases. The $m_{\rm max}$ posterior for the \textsc{peak} model is almost identical to the \textsc{bimodal} model.

Our estimate of the Galactic NS population's maximum mass comes from the double-Gaussian model of \citet{FarrChatziioannou2020}, which predicts $m_{\rm max} = 2.3^{+0.8}_{-0.3}\Msun$. The corresponding posterior distribution is displayed in Fig.~\ref{fig:mmax}. As can be seen, this Galactic maximum mass is completely consistent with the \textsc{flat} and \textsc{bimodal} estimates of $m_{\rm max}$ within current uncertainties: there is no discrepancy in the maximum masses of Galactic and extragalactic NSs.

For the maximum TOV mass---the maximum mass of a nonrotating NS---predicted by the dense-matter EOS, we take the value of $M_{\rm TOV} = 2.2^{+0.4}_{-0.2}\Msun$ obtained by \citet{LandryEssick2020} from nonparametric inference based on radio pulsar, X-ray and GW observations of NSs, an analysis that did not interpret GW190814's secondary as a NS. We find good agreement between the EOS-informed maximum NS mass posterior, also shown in Fig.~\ref{fig:mmax}, and that inferred from the GW population with both the \textsc{flat} and \textsc{bimodal} models: the maximum mass for extragalactic NSs is consistent with $M_{\rm TOV}$, suggesting that stellar and binary evolution are able to produce GW binaries containing the heaviest NSs. This conclusion persists even when GW190814 is incorporated into the population inference as an NSBH observation, as long we also account for it in the analysis of the EOS; as a high-mass outlier, it dictates $M_{\rm TOV}$.

\begin{figure}[tb!]
\centering
\includegraphics[width=0.98\columnwidth,trim={0 0 0 0},clip]{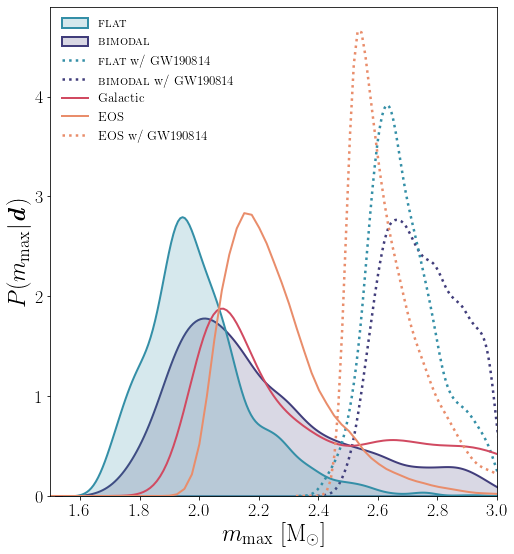}
\caption{Inferred maximum mass for the population of NSs in GW binaries. Posterior distributions $P(m_{\rm max} | \boldsymbol{d})$ for the maximum mass parameter in the \textsc{flat} and \textsc{bimodal} models are compared against the maximum mass inferred from the Galactic NS population in \citet{FarrChatziioannou2020} and the maximum TOV mass from the EOS inference in \citet{LandryEssick2020}. The dotted curves show the maximum mass posterior when GW190814's secondary is treated as a NS; we simulate the impact of this assumption on $M_{\rm TOV}$ by discarding samples with $M_{\rm TOV} < 2.5\Msun$, the 90\% credible lower bound on GW190814's secondary mass, from the \citet{LandryEssick2020} posterior.
\figlabel{fig:mmax}}
\end{figure}

\paragraph{GW190814}

The results presented up to this point have relied on four GW observations---GW170817, GW190425, GW200105, and GW200115---that are readily interpretable as BNS or NSBH mergers. In this subsection, we also consider GW190814, whose secondary component has a mass of $2.6^{+0.1}_{-0.1}\Msun$~\citep{GW190814}. Although its nature is uncertain \citep[e.g.][]{MostPapenfort2020,BiswasNandi2021,TewsPang2021}, and comparisons with various maximum NS mass estimates suggest its secondary may be a low-mass BH~\citep{GW190814,EssickLandry2020}, we now entertain the scenario in which it is a NS and revisit our population inference under this assumption. Because GW190814 is such an outlier relative to the other GW observations, its inclusion in the population has a significant impact on the inferred mass distribution.

Its primary effect is to flatten and extend the NS mass distribution beyond $2.5\Msun$, as can be seen in Figs.~\ref{fig:flatmass}, \ref{fig:galactic} and \ref{fig:peak} when it is included in the \textsc{flat}, \textsc{bimodal} and \textsc{peak} analyses, respectively. Indeed, the $m_{\rm max}$ posterior is particularly sensitive to the inclusion of GW190814. For the \textsc{flat} model, the inferred maximum mass in the GW population of NSs shifts from $2.0^{+0.4}_{-0.3}\Msun$ to $2.7^{+0.2}_{-0.2}\Msun$, as illustrated in Fig.~\ref{fig:mmax}. For the \textsc{bimodal} model, a comparable shift from $2.1^{+0.7}_{-0.3}\Msun$ to $2.7^{+0.3}_{-0.1}\Msun$ occurs (the $m_{\rm max}$ values for the \textsc{peak} model are very similar). If the population of NSs in GW binaries is taken to include GW190814's secondary component, there is likely a difference between the maximum masses observed in the Galactic and GW populations, although the uncertainties in $m_{\rm max}$ are large enough that the posteriors still overlap. This difference could be indicative of a radio selection effect like the one discussed in~\citet{GalaudageAdamcewicz2021}. Of course, the simplest reconciliation of this discrepancy between the Galactic and GW populations is that GW190814 does not contain a NS, as is suggested by a detailed comparison of its secondary mass with an EOS-informed estimate of $M_{\rm TOV}$ based on other astrophysical observations~\citep{EssickLandry2020}.

\paragraph{Pairing function}

Thus far, our population models have assumed random pairing of NSs into BNSs. We now revise that assumption and investigate mass-ratio--dependent pairing as an alternative, using a power-law pairing function as in \citet{FishbachHolz2020}. We choose $\beta = 2$ as our fixed BNS pairing parameter to match our aforementioned alternative NSBH pairing parameter choice $\beta_{\textsc{bh}} = 2$. We introduce \textsc{flat\_q}, \textsc{peak\_q} and \textsc{bimodal\_q} models that differ from the fiducial ones only by this choice of $\beta$ in Eq.~\eqref{bns_mdistr}. For each of these new models, we repeat the analysis described at the beginning of this section and examine how the inferred mass distributions and population parameter constraints change. 
We then compute AICs to determine whether the data prefer random or $q$-dependent pairing.

The medians of the inferred mass distributions for the \textsc{flat\_q}, \textsc{bimodal\_q} and \textsc{peak\_q} models are plotted in Figs.~\ref{fig:flatmass}, \ref{fig:galactic} and \ref{fig:peak}, respectively, alongside their random-pairing counterparts. In all cases there is little difference between the mass distributions inferred with and without $q$-dependent pairing. This is also apparent from the population parameter constraints in Table~\ref{tab:metrics}.

We compute AICs for the new models, listing their $\Delta {\rm AIC}$ values relative to the \textsc{flat} model in Table~\ref{tab:metrics}. 
The AIC is marginally larger for the \textsc{flat\_q}, \textsc{peak\_q} and \textsc{bimodal\_q} models than their respective random-pairing counterparts. Overall, we find no systematic evidence that a $q^2$ pairing function is preferred over random pairing for BNSs. This conclusion also holds when we test a more extreme pairing power law ($\beta = 7$); GW observations cannot yet constrain the BNS pairing function.

\section{Conclusions}
With this first look at NSs in GW binaries as a population, we find a broad spread of masses over the range compatible with NSs, consistent with a uniform mass distribution. There is no clear evidence in the GW population of the double-peaked structure inferred from Galactic NS mass measurements, nor of the narrow low-mass peak associated specifically with Galactic double NSs. Indeed, we find that high-mass NSs are relatively more prevalent in GW binaries than in the those surveyed with electromagnetic astronomy. However, the maximum NS mass we infer from the GW population agrees with the sharp high-mass cutoff in the Galactic NS mass distribution, provided we do not interpret the outlier event GW190814 as an NSBH. Regardless of GW190814's classification, this maximum mass is consistent with the maximum nonrotating NS mass supported by the dense-matter EOS. The current set of extragalactic NS observations does not allow us to distinguish whether NSs pair randomly or preferentially in equal-mass BNSs. Given the small number of observations currently at hand, near-future NS  mass  measurements with GWs will have a significant impact on this early picture of the  population. \\

The authors thank Katerina Chatziioannou, Reed Essick, Maya Fishbach, Shanika Galaudage, Richard O'Shaughnessy, Simon Stevenson and the LIGO-Virgo Rates \& Populations working group for useful feedback about the manuscript. P.L. is supported by National Science Foundation award PHY-1836734 and by a gift from the Dan Black Family Foundation to the Nicholas \& Lee Begovich Center for Gravitational-Wave Physics \& Astronomy. J.R. thanks NSF PHY-1836734 and PHY-1806962 and the Carnegie Observatories for support. This research makes use of data,  software and/or web  tools  obtained  from  the  Gravitational  Wave  Open Science Center (\url{https://www.gw-openscience.org}), a service of LIGO Laboratory,  the LIGO Scientific Collaboration and the Virgo Collaboration. LIGO is funded by the U.S. National Science Foundation.  Virgo is funded by the French Centre National de Recherche Scientifique (CNRS), the Italian Istituto Nazionale della Fisica Nucleare (INFN) and the Dutch Nikhef, with contributions by Polish and Hungarian institutes. The authors are grateful for computational resources provided by the LIGO Laboratory and supported by National Science Foundation grants PHY-0757058 and PHY-0823459. This material is based upon work supported by NSF's LIGO Laboratory, which is a major facility fully funded by the National Science Foundation.

\pagebreak

\bibliography{references}

\begin{thebibliography}{}
\expandafter\ifx\csname natexlab\endcsname\relax\def\natexlab#1{#1}\fi
\providecommand{\url}[1]{\href{#1}{#1}}
\providecommand{\dodoi}[1]{doi:~\href{http://doi.org/#1}{\nolinkurl{#1}}}
\providecommand{\doeprint}[1]{\href{http://ascl.net/#1}{\nolinkurl{http://ascl.net/#1}}}
\providecommand{\doarXiv}[1]{\href{https://arxiv.org/abs/#1}{\nolinkurl{https://arxiv.org/abs/#1}}}

\bibitem[{{Abbott} {et~al.}(2016){Abbott}, {Abbott}, {Abbott}, {Abernathy},
  {Acernese}, {Ackley}, {Adams}, {Adams}, {Addesso}, {Adhikari},
  {et~al.}}]{2016ApJ...832L..21A}
{Abbott}, B.~P., {Abbott}, R., {Abbott}, T.~D., {et~al.} 2016, \apjl, 832, L21,
  \dodoi{10.3847/2041-8205/832/2/L21}

\bibitem[{{Abbott} {et~al.}(2017){Abbott}, {Abbott}, {Abbott}, {Acernese},
  {Ackley}, {Adams}, {Adams}, {Addesso}, {Adhikari}, {Adya}, \&
  et~al.}]{GW170817}
---. 2017, \prl, 119, 161101, \dodoi{10.1103/PhysRevLett.119.161101}

\bibitem[{{Abbott} {et~al.}(2019){Abbott}, {Abbott}, {Abbott}, {Abraham},
  {Acernese}, {Ackley}, {Adams}, {Adhikari}, {Adya}, {Affeldt}, \&
  et~al.}]{O1O2pop}
---. 2019, ApJL, 882, L24, \dodoi{10.3847/2041-8213/ab3800}

\bibitem[{{Abbott} {et~al.}(2020{\natexlab{a}}){Abbott}, {Abbott}, {Abbott},
  {Abraham}, {Acernese}, {Ackley}, {Adams}, {Adhikari}, {Adya}, {Affeldt}, \&
  et~al.}]{GW190425}
---. 2020{\natexlab{a}}, \apjl, 892, L3, \dodoi{10.3847/2041-8213/ab75f5}

\bibitem[{{Abbott} {et~al.}(2020{\natexlab{b}}){Abbott}, {Abbott}, {Abraham},
  {Acernese}, {Ackley}, {Adams}, {Adhikari}, {Adya}, {Affeldt}, {Agathos}, \&
  et~al.}]{GW190814}
{Abbott}, R., {Abbott}, T.~D., {Abraham}, S., {et~al.} 2020{\natexlab{b}},
  ApJL, 896, L44, \dodoi{10.3847/2041-8213/ab960f}

\bibitem[{{Abbott} {et~al.}(2021{\natexlab{a}}){Abbott}, {Abbott}, {Abraham},
  {Acernese}, {Ackley}, {Adams}, {Adams}, {Adhikari}, {Adya}, {Affeldt}, \&
  et~al.}]{NSBHs}
---. 2021{\natexlab{a}}, ApJL, 915, L5, \dodoi{10.3847/2041-8213/ac082e}

\bibitem[{{Abbott} {et~al.}(2021{\natexlab{b}}){Abbott}, {Abbott}, {Abraham},
  {Acernese}, {Ackley}, {Adams}, {Adams}, {Adhikari}, {Adya}, {Affeldt}, \&
  et~al.}]{O3apop}
---. 2021{\natexlab{b}}, ApJL, 913, L7, \dodoi{10.3847/2041-8213/abe949}

\bibitem[{{Abbott} {et~al.}(2021{\natexlab{c}}){Abbott}, {Abbott}, {Abraham},
  {Acernese}, {Ackley}, {Adams}, {Adhikari}, {Adya}, {Affeldt}, {Agathos},
  {Agatsuma}, {Aggarwal}, {Aguiar}, {Aich}, {Aiello}, {Ain}, {Ajith}, {Allen},
  {Allocca}, {Altin}, {Amato}, {Anand}, {Ananyeva}, {Anderson}, {Anderson},
  {Angelova}, {Ansoldi}, {Antier}, {Appert}, {Arai}, {Araya}, {Areeda},
  {Ar{\`e}ne}, {Arnaud}, {Aronson}, {Arun}, {Ascenzi}, {Ashton}, {Aston},
  {Astone}, {Aubin}, {Aufmuth}, {AultONeal}, {Austin}, {Avendano}, {Babak},
  {Bacon}, {Badaracco}, {Bader}, {Bae}, {Baer}, {Baird}, {Baldaccini},
  {Ballardin}, {Ballmer}, {Bals}, {Balsamo}, {Baltus}, {Banagiri}, {Bankar},
  {Bankar}, {Barayoga}, {Barbieri}, {Barish}, {Barker}, {Barkett}, {Barneo},
  {Barone}, {Barr}, {Barsotti}, {Barsuglia}, {Barta}, {Bartlett}, {Bartos},
  {Bassiri}, {Basti}, {Bawaj}, {Bayley}, {Bazzan}, {B{\'e}csy}, {Bejger},
  {Belahcene}, {Bell}, {Beniwal}, {Benjamin}, {Bentley}, {Bergamin}, {Berger},
  {Bergmann}, {Bernuzzi}, {Berry}, {Bersanetti}, {Bertolini}, {Betzwieser},
  {Bhandare}, {Bhandari}, {Bidler}, {Biggs}, {Bilenko}, {Billingsley},
  {Birney}, {Birnholtz}, {Biscans}, {Bischi}, {Biscoveanu}, {Bisht},
  {Bissenbayeva}, {Bitossi}, {Bizouard}, {Blackburn}, {Blackman}, {Blair},
  {Blair}, {Blair}, {Bobba}, {Bode}, {Boer}, {Boetzel}, {Bogaert}, {Bondu},
  {Bonilla}, {Bonnand}, {Booker}, {Boom}, {Bork}, {Boschi}, {Bose},
  {Bossilkov}, {Bosveld}, {Bouffanais}, {Bozzi}, {Bradaschia}, {Brady},
  {Bramley}, {Branchesi}, {Brau}, {Breschi}, {Briant}, {Briggs}, {Brighenti},
  {Brillet}, {Brinkmann}, {Brockill}, {Brooks}, {Brooks}, {Brown}, {Brunett},
  {Bruno}, {Bruntz}, {Buikema}, {Bulik}, {Bulten}, {Buonanno}, {Buskulic},
  {Byer}, {Cabero}, {Cadonati}, {Cagnoli}, {Cahillane}, {Bustillo},
  {Callaghan}, {Callister}, {Calloni}, {Camp}, {Canepa}, {Cannon}, {Cao},
  {Cao}, {Carapella}, {Carbognani}, {Caride}, {Carney}, {Carullo}, {Diaz},
  {Casentini}, {Casta{\~n}eda}, {Caudill}, {Cavagli{\`a}}, {Cavalier},
  {Cavalieri}, {Cella}, {Cerd{\'a}-Dur{\'a}n}, {Cesarini}, {Chaibi},
  {Chakravarti}, {Chan}, {Chan}, {Chao}, {Charlton}, {Chase},
  {Chassande-Mottin}, {Chatterjee}, {Chaturvedi}, {Chen}, {Chen}, {Chen},
  {Cheng}, {Cheong}, {Chia}, {Chiadini}, {Chierici}, {Chincarini}, {Chiummo},
  {Cho}, {Cho}, {Cho}, {Christensen}, {Chu}, {Chua}, {Chung}, {Chung}, {Ciani},
  {Ciecielag}, {Cie{\'s}lar}, {Ciobanu}, {Ciolfi}, {Cipriano}, {Cirone},
  {Clara}, {Clark}, {Clearwater}, {Clesse}, {Cleva}, {Coccia}, {Cohadon},
  {Cohen}, {Colleoni}, {Collette}, {Collins}, {Colpi}, {Constancio}, {Conti},
  {Cooper}, {Corban}, {Corbitt}, {Cordero-Carri{\'o}n}, {Corezzi}, {Corley},
  {Cornish}, {Corre}, {Corsi}, {Cortese}, {Costa}, {Cotesta}, {Coughlin},
  {Coughlin}, {Coulon}, {Countryman}, {Couvares}, {Covas}, {Coward}, {Cowart},
  {Coyne}, {Coyne}, {Creighton}, {Creighton}, {Cripe}, {Croquette}, {Crowder},
  {Cudell}, {Cullen}, {Cumming}, {Cummings}, {Cunningham}, {Cuoco}, {Curylo},
  {Canton}, {D{\'a}lya}, {Dana}, {Daneshgaran-Bajastani}, {D'Angelo},
  {Danilishin}, {D'Antonio}, {Danzmann}, {Darsow-Fromm}, {Dasgupta}, {Datrier},
  {Dattilo}, {Dave}, {Davier}, {Davies}, {Davis}, {Daw}, {DeBra},
  {Deenadayalan}, {Degallaix}, {De Laurentis}, {Del{\'e}glise}, {Delfavero},
  {De Lillo}, {Del Pozzo}, {DeMarchi}, {D'Emilio}, {Demos}, {Dent}, {De
  Pietri}, {De Rosa}, {De Rossi}, {DeSalvo}, {de Varona}, {Dhurandhar},
  {D{\'\i}az}, {Diaz-Ortiz}, {Dietrich}, {Di Fiore}, {Di Fronzo}, {Di Giorgio},
  {Di Giovanni}, {Di Giovanni}, {Di Girolamo}, {Di Lieto}, {Ding}, {Di Pace},
  {Di Palma}, {Di Renzo}, {Divakarla}, {Dmitriev}, {Doctor}, {Donovan},
  {Dooley}, {Doravari}, {Dorrington}, {Downes}, {Drago}, {Driggers}, {Du},
  {Ducoin}, {Dupej}, {Durante}, {D'Urso}, {Dwyer}, {Easter}, {Eddolls},
  {Edelman}, {Edo}, {Edy}, {Effler}, {Ehrens}, {Eichholz}, {Eikenberry},
  {Eisenmann}, {Eisenstein}, {Ejlli}, {Errico}, {Essick}, {Estelles},
  {Estevez}, {Etienne}, {Etzel}, {Evans}, {Evans}, {Ewing}, {Fafone},
  {Fairhurst}, {Fan}, {Farinon}, {Farr}, {Farr}, {Fauchon-Jones}, {Favata},
  {Fays}, {Fazio}, {Feicht}, {Fejer}, {Feng}, {Fenyvesi}, {Ferguson},
  {Fernandez-Galiana}, {Ferrante}, {Ferreira}, {Ferreira}, {Fidecaro}, {Fiori},
  {Fiorucci}, {Fishbach}, {Fisher}, {Fittipaldi}, {Fitz-Axen}, {Fiumara},
  {Flaminio}, {Floden}, {Flynn}, {Fong}, {Font}, {Forsyth}, {Fournier},
  {Frasca}, {Frasconi}, {Frei}, {Freise}, {Frey}, {Frey}, {Fritschel},
  {Frolov}, {Fronz{\`e}}, {Fulda}, {Fyffe}, {Gabbard}, {Gadre}, {Gaebel},
  {Gair}, {Galaudage}, {Ganapathy}, {Gaonkar}, {Garc{\'\i}a-Quir{\'o}s},
  {Garufi}, {Gateley}, {Gaudio}, {Gayathri}, {Gemme}, {Genin}, {Gennai},
  {George}, {George}, {Gergely}, {Ghonge}, {Ghosh}, {Ghosh}, {Ghosh},
  {Giacomazzo}, {Giaime}, {Giardina}, {Gibson}, {Gier}, {Gill}, {Glanzer},
  {Gniesmer}, {Godwin}, {Goetz}, {Goetz}, {Gohlke}, {Goncharov},
  {Gonz{\'a}lez}, {Gopakumar}, {Gossan}, {Gosselin}, {Gouaty}, {Grace},
  {Grado}, {Granata}, {Grant}, {Gras}, {Grassia}, {Gray}, {Gray}, {Greco},
  {Green}, {Green}, {Gretarsson}, {Griggs}, {Grignani}, {Grimaldi}, {Grimm},
  {Grote}, {Grunewald}, {Gruning}, {Guidi}, {Guimaraes}, {Guix{\'e}}, {Gulati},
  {Guo}, {Gupta}, {Gupta}, {Gupta}, {Gustafson}, {Gustafson}, {Haegel},
  {Halim}, {Hall}, {Hamilton}, {Hammond}, {Haney}, {Hanke}, {Hanks}, {Hanna},
  {Hannam}, {Hannuksela}, {Hansen}, {Hanson}, {Harder}, {Hardwick}, {Haris},
  {Harms}, {Harry}, {Harry}, {Hasskew}, {Haster}, {Haughian}, {Hayes}, {Healy},
  {Heidmann}, {Heintze}, {Heinze}, {Heitmann}, {Hellman}, {Hello}, {Hemming},
  {Hendry}, {Heng}, {Hennes}, {Hennig}, {Heurs}, {Hild}, {Hinderer}, {Hoback},
  {Hochheim}, {Hofgard}, {Hofman}, {Holgado}, {Holland}, {Holt}, {Holz},
  {Hopkins}, {Horst}, {Hough}, {Howell}, {Hoy}, {Huang}, {H{\"u}bner},
  {Huerta}, {Huet}, {Hughey}, {Hui}, {Husa}, {Huttner}, {Huxford},
  {Huynh-Dinh}, {Idzkowski}, {Iess}, {Inchauspe}, {Ingram}, {Intini}, {Isac},
  {Isi}, {Iyer}, {Jacqmin}, {Jadhav}, {Jadhav}, {James}, {Jani}, {Janthalur},
  {Jaranowski}, {Jariwala}, {Jaume}, {Jenkins}, {Jiang}, {Johns}, {Jones},
  {Jones}, {Jones}, {Jones}, {Jones}, {Jonker}, {Ju}, {Junker}, {Kalaghatgi},
  {Kalogera}, {Kamai}, {Kandhasamy}, {Kang}, {Kanner}, {Kapadia}, {Karki},
  {Kashyap}, {Kasprzack}, {Kastaun}, {Katsanevas}, {Katsavounidis}, {Katzman},
  {Kaufer}, {Kawabe}, {K{\'e}f{\'e}lian}, {Keitel}, {Keivani}, {Kennedy},
  {Key}, {Khadka}, {Khalili}, {Khan}, {Khan}, {Khan}, {Khazanov}, {Khetan},
  {Khursheed}, {Kijbunchoo}, {Kim}, {Kim}, {Kim}, {Kim}, {Kim}, {Kim}, {Kim},
  {Kimball}, {King}, {Kinley-Hanlon}, {Kirchhoff}, {Kissel}, {Kleybolte},
  {Klimenko}, {Knowles}, {Koch}, {Koehlenbeck}, {Koekoek}, {Koley},
  {Kondrashov}, {Kontos}, {Koper}, {Korobko}, {Korth}, {Kovalam}, {Kozak},
  {Kringel}, {Krishnendu}, {Kr{\'o}lak}, {Krupinski}, {Kuehn}, {Kumar},
  {Kumar}, {Kumar}, {Kumar}, {Kumar}, {Kuo}, {Kutynia}, {Lackey}, {Laghi},
  {Lalande}, {Lam}, {Lamberts}, {Landry}, {Lane}, {Lang}, {Lange}, {Lantz},
  {Lanza}, {La Rosa}, {Lartaux-Vollard}, {Lasky}, {Laxen}, {Lazzarini},
  {Lazzaro}, {Leaci}, {Leavey}, {Lecoeuche}, {Lee}, {Lee}, {Lee}, {Lee}, {Lee},
  {Lehmann}, {Leroy}, {Letendre}, {Levin}, {Li}, {Li}, {li}, {Li}, {Li},
  {Linde}, {Linker}, {Linley}, {Littenberg}, {Liu}, {Liu},
  {Llorens-Monteagudo}, {Lo}, {Lockwood}, {London}, {Longo}, {Lorenzini},
  {Loriette}, {Lormand}, {Losurdo}, {Lough}, {Lousto}, {Lovelace}, {L{\"u}ck},
  {Lumaca}, {Lundgren}, {Ma}, {Macas}, {Macfoy}, {MacInnis}, {Macleod},
  {MacMillan}, {Macquet}, {Hernandez}, {Maga{\~n}a-Sandoval}, {Magee},
  {Majorana}, {Maksimovic}, {Malik}, {Man}, {Mandic}, {Mangano}, {Mansell},
  {Manske}, {Mantovani}, {Mapelli}, {Marchesoni}, {Marion}, {M{\'a}rka},
  {M{\'a}rka}, {Markakis}, {Markosyan}, {Markowitz}, {Maros}, {Marquina},
  {Marsat}, {Martelli}, {Martin}, {Martin}, {Martinez}, {Martynov},
  {Masalehdan}, {Mason}, {Massera}, {Masserot}, {Massinger}, {Masso-Reid},
  {Mastrogiovanni}, {Matas}, {Matichard}, {Mavalvala}, {Maynard}, {McCann},
  {McCarthy}, {McClelland}, {McCormick}, {McCuller}, {McGuire}, {McIsaac},
  {McIver}, {McManus}, {McRae}, {McWilliams}, {Meacher}, {Meadors}, {Mehmet},
  {Mehta}, {Villa}, {Melatos}, {Mendell}, {Mercer}, {Mereni}, {Merfeld},
  {Merilh}, {Merritt}, {Merzougui}, {Meshkov}, {Messenger}, {Messick},
  {Metzdorff}, {Meyers}, {Meylahn}, {Mhaske}, {Miani}, {Miao}, {Michaloliakos},
  {Michel}, {Middleton}, {Milano}, {Miller}, {Millhouse}, {Mills}, {Milotti},
  {Milovich-Goff}, {Minazzoli}, {Minenkov}, {Mishkin}, {Mishra}, {Mistry},
  {Mitra}, {Mitrofanov}, {Mitselmakher}, {Mittleman}, {Mo}, {Mogushi},
  {Mohapatra}, {Mohite}, {Molina-Ruiz}, {Mondin}, {Montani}, {Moore}, {Moraru},
  {Morawski}, {Moreno}, {Morisaki}, {Mours}, {Mow-Lowry}, {Mozzon},
  {Muciaccia}, {Mukherjee}, {Mukherjee}, {Mukherjee}, {Mukherjee}, {Mukund},
  {Mullavey}, {Munch}, {Mu{\~n}iz}, {Murray}, {Nagar}, {Nardecchia},
  {Naticchioni}, {Nayak}, {Neil}, {Neilson}, {Nelemans}, {Nelson}, {Nery},
  {Neunzert}, {Ng}, {Ng}, {Nguyen}, {Nguyen}, {Nichols}, {Nichols}, {Nissanke},
  {Nocera}, {Noh}, {North}, {Nothard}, {Nuttall}, {Oberling}, {O'Brien},
  {Oganesyan}, {Ogin}, {Oh}, {Oh}, {Ohme}, {Ohta}, {Okada}, {Oliver},
  {Olivetto}, {Oppermann}, {Oram}, {O'Reilly}, {Ormiston}, {Ortega},
  {O'Shaughnessy}, {Ossokine}, {Osthelder}, {Ottaway}, {Overmier}, {Owen},
  {Pace}, {Pagano}, {Page}, {Pagliaroli}, {Pai}, {Pai}, {Palamos}, {Palashov},
  {Palomba}, {Pan}, {Panda}, {Pang}, {Pankow}, {Pannarale}, {Pant}, {Paoletti},
  {Paoli}, {Parida}, {Parker}, {Pascucci}, {Pasqualetti}, {Passaquieti},
  {Passuello}, {Patricelli}, {Payne}, {Pearlstone}, {Pechsiri}, {Pedersen},
  {Pedraza}, {Pele}, {Penn}, {Perego}, {Perez}, {P{\'e}rigois}, {Perreca},
  {Perri{\`e}s}, {Petermann}, {Pfeiffer}, {Phelps}, {Phukon}, {Piccinni},
  {Pichot}, {Piendibene}, {Piergiovanni}, {Pierro}, {Pillant}, {Pinard},
  {Pinto}, {Piotrzkowski}, {Pirello}, {Pitkin}, {Plastino}, {Poggiani}, {Pong},
  {Ponrathnam}, {Popolizio}, {Porter}, {Powell}, {Prajapati}, {Prasai},
  {Prasanna}, {Pratten}, {Prestegard}, {Principe}, {Prodi}, {Prokhorov},
  {Punturo}, {Puppo}, {P{\"u}rrer}, {Qi}, {Quetschke}, {Quinonez}, {Raab},
  {Raaijmakers}, {Radkins}, {Radulesco}, {Raffai}, {Rafferty}, {Raja}, {Rajan},
  {Rajbhandari}, {Rakhmanov}, {Ramirez}, {Ramos-Buades}, {Rana}, {Rao},
  {Rapagnani}, {Raymond}, {Razzano}, {Read}, {Regimbau}, {Rei}, {Reid},
  {Reitze}, {Rettegno}, {Ricci}, {Richardson}, {Richardson}, {Ricker},
  {Riemenschneider}, {Riles}, {Rizzo}, {Robertson}, {Robinet}, {Rocchi},
  {Rodriguez-Soto}, {Rolland}, {Rollins}, {Roma}, {Romanelli}, {Romano},
  {Romel}, {Romero-Shaw}, {Romie}, {Rose}, {Rose}, {Rose}, {Rosi{\'n}ska},
  {Rosofsky}, {Ross}, {Rowan}, {Rowlinson}, {Roy}, {Roy}, {Roy}, {Ruggi},
  {Rutins}, {Ryan}, {Sachdev}, {Sadecki}, {Sakellariadou}, {Salafia},
  {Salconi}, {Saleem}, {Samajdar}, {Sanchez}, {Sanchez}, {Sanchis-Gual},
  {Sanders}, {Santiago}, {Santos}, {Sarin}, {Sassolas}, {Sathyaprakash},
  {Sauter}, {Savage}, {Savant}, {Sawant}, {Sayah}, {Schaetzl}, {Schale},
  {Scheel}, {Scheuer}, {Schmidt}, {Schnabel}, {Schofield}, {Sch{\"o}nbeck},
  {Schreiber}, {Schulte}, {Schutz}, {Schwarm}, {Schwartz}, {Scott}, {Scott},
  {Seidel}, {Sellers}, {Sengupta}, {Sennett}, {Sentenac}, {Sequino}, {Sergeev},
  {Setyawati}, {Shaddock}, {Shaffer}, {Shahriar}, {Sharma}, {Sharma},
  {Shawhan}, {Shen}, {Shikauchi}, {Shink}, {Shoemaker}, {Shoemaker}, {Shukla},
  {ShyamSundar}, {Siellez}, {Sieniawska}, {Sigg}, {Singer}, {Singh}, {Singh},
  {Singha}, {Singhal}, {Sintes}, {Sipala}, {Skliris}, {Slagmolen},
  {Slaven-Blair}, {Smetana}, {Smith}, {Smith}, {Somala}, {Son}, {Soni},
  {Sorazu}, {Sordini}, {Sorrentino}, {Souradeep}, {Sowell}, {Spencer}, {Spera},
  {Srivastava}, {Srivastava}, {Staats}, {Stachie}, {Standke}, {Steer},
  {Steinke}, {Steinlechner}, {Steinlechner}, {Steinmeyer}, {Stocks}, {Stops},
  {Stover}, {Strain}, {Stratta}, {Strunk}, {Sturani}, {Stuver}, {Sudhagar},
  {Sudhir}, {Summerscales}, {Sun}, {Sunil}, {Sur}, {Suresh}, {Sutton},
  {Swinkels}, {Szczepa{\'n}czyk}, {Tacca}, {Tait}, {Talbot}, {Tanasijczuk},
  {Tanner}, {Tao}, {T{\'a}pai}, {Tapia}, {Martin}, {Tasson}, {Taylor},
  {Tenorio}, {Terkowski}, {Thirugnanasambandam}, {Thomas}, {Thomas},
  {Thompson}, {Thondapu}, {Thorne}, {Thrane}, {Tinsman}, {Saravanan}, {Tiwari},
  {Tiwari}, {Tiwari}, {Toland}, {Tonelli}, {Tornasi}, {Torres-Forn{\'e}},
  {Torrie}, {Tosta e Melo}, {T{\"o}yr{\"a}}, {Trail}, {Travasso}, {Traylor},
  {Tringali}, {Tripathee}, {Trovato}, {Trudeau}, {Tsang}, {Tse}, {Tso},
  {Tsukada}, {Tsuna}, {Tsutsui}, {Turconi}, {Ubhi}, {Ueno}, {Ugolini},
  {Unnikrishnan}, {Urban}, {Usman}, {Utina}, {Vahlbruch}, {Vajente}, {Valdes},
  {Valentini}, {Vallisneri}, {van Bakel}, {van Beuzekom}, {van den Brand}, {Van
  Den Broeck}, {Vander-Hyde}, {van der Schaaf}, {Van Heijningen}, {van Veggel},
  {Vardaro}, {Varma}, {Vass}, {Vas{\'u}th}, {Vecchio}, {Vedovato}, {Veitch},
  {Veitch}, {Venkateswara}, {Venugopalan}, {Verkindt}, {Veske}, {Vetrano},
  {Vicer{\'e}}, {Viets}, {Vinciguerra}, {Vine}, {Vinet}, {Vitale}, {Vivanco},
  {Vo}, {Vocca}, {Vorvick}, {Vyatchanin}, {Wade}, {Wade}, {Wade}, {Walet},
  {Walker}, {Wallace}, {Wallace}, {Walsh}, {Wang}, {Wang}, {Wang}, {Wang},
  {Ward}, {Warden}, {Warner}, {Was}, {Watchi}, {Weaver}, {Wei}, {Weinert},
  {Weinstein}, {Weiss}, {Wellmann}, {Wen}, {We{\ss}els}, {Westhouse}, {Wette},
  {Whelan}, {Whiting}, {Whittle}, {Wilken}, {Williams}, {Williams},
  {Williamson}, {Willis}, {Willke}, {Winkler}, {Wipf}, {Wittel}, {Woan},
  {Woehler}, {Wofford}, {Wong}, {Wright}, {Wu}, {Wysocki}, {Xiao}, {Yamamoto},
  {Yang}, {Yang}, {Yang}, {Yap}, {Yazback}, {Yeeles}, {Yu}, {Yu}, {Yuen},
  {Zadro{\.z}ny}, {Zadro{\.z}ny}, {Zanolin}, {Zelenova}, {Zendri}, {Zevin},
  {Zhang}, {Zhang}, {Zhang}, {Zhao}, {Zhao}, {Zhou}, {Zhou}, {Zhu},
  {Zimmerman}, {Zucker}, {Zweizig}, \& {LIGO Scientific
  Collaboration}}]{2021SoftX..1300658A}
---. 2021{\natexlab{c}}, SoftX, 13, 100658, \dodoi{10.1016/j.softx.2021.100658}

\bibitem[{{Abbott} {et~al.}(2021{\natexlab{d}}){Abbott}, {Abbott}, {Abraham},
  {Acernese}, {Ackley}, {Adams}, {Adams}, {Adhikari}, {Adya}, {Affeldt}, \&
  et~al.}]{GWTC2}
---. 2021{\natexlab{d}}, PhRvX, 11, 021053, \dodoi{10.1103/PhysRevX.11.021053}

\bibitem[{{Acernese} {et~al.}(2015){Acernese}, {Agathos}, {Agatsuma}, {Aisa},
  {Allemandou}, {Allocca}, {Amarni}, {Astone}, {Balestri}, {Ballardin}, \&
  et~al.}]{aVirgo}
{Acernese}, F., {Agathos}, M., {Agatsuma}, K., {et~al.} 2015, CQG, 32, 024001,
  \dodoi{10.1088/0264-9381/32/2/024001}

\bibitem[{Akaike(1981)}]{AKAIKE19813}
Akaike, H. 1981, Journal of Econometrics, 16, 3,
  \dodoi{10.1016/0304-4076(81)90071-3}

\bibitem[{{Alsing} {et~al.}(2018){Alsing}, {Silva}, \&
  {Berti}}]{AlsingSilva2018}
{Alsing}, J., {Silva}, H.~O., \& {Berti}, E. 2018, \mnras, 478, 1377,
  \dodoi{10.1093/mnras/sty1065}

\bibitem[{{Antoniadis} {et~al.}(2016){Antoniadis}, {Tauris}, {Ozel}, {Barr},
  {Champion}, \& {Freire}}]{2016arXiv160501665A}
{Antoniadis}, J., {Tauris}, T.~M., {Ozel}, F., {et~al.} 2016, arXiv:1605.01665.
\newblock \doarXiv{1605.01665}

\bibitem[{{Biswas} {et~al.}(2021){Biswas}, {Nandi}, {Char}, {Bose}, \&
  {Stergioulas}}]{BiswasNandi2021}
{Biswas}, B., {Nandi}, R., {Char}, P., {Bose}, S., \& {Stergioulas}, N. 2021,
  MNRAS, \dodoi{10.1093/mnras/stab1383}

\bibitem[{{Broekgaarden} {et~al.}(2021){Broekgaarden}, {Berger}, {Neijssel},
  {Vigna-G{\'o}mez}, {Chattopadhyay}, {Stevenson}, {Chruslinska}, {Justham},
  {de Mink}, \& {Mandel}}]{2021arXiv210302608B}
{Broekgaarden}, F.~S., {Berger}, E., {Neijssel}, C.~J., {et~al.} 2021,
  MNRAS.tmp, \dodoi{10.1093/mnras/stab2716}

\bibitem[{{Chattopadhyay} {et~al.}(2020){Chattopadhyay}, {Stevenson}, {Hurley},
  {Rossi}, \& {Flynn}}]{2020MNRAS.494.1587C}
{Chattopadhyay}, D., {Stevenson}, S., {Hurley}, J.~R., {Rossi}, L.~J., \&
  {Flynn}, C. 2020, \mnras, 494, 1587, \dodoi{10.1093/mnras/staa756}

\bibitem[{{Chatziioannou} \& {Farr}(2020)}]{ChatziioannouFarr2020}
{Chatziioannou}, K., \& {Farr}, W.~M. 2020, PhRvD, 102, 064063,
  \dodoi{10.1103/PhysRevD.102.064063}

\bibitem[{{Cromartie} {et~al.}(2020){Cromartie}, {Fonseca}, {Ransom},
  {Demorest}, {Arzoumanian}, {Blumer}, {Brook}, {DeCesar}, {Dolch}, {Ellis},
  {Ferdman}, {Ferrara}, {Garver-Daniels}, {Gentile}, {Jones}, {Lam}, {Lorimer},
  {Lynch}, {McLaughlin}, {Ng}, {Nice}, {Pennucci}, {Spiewak}, {Stairs},
  {Stovall}, {Swiggum}, \& {Zhu}}]{2020NatAs...4...72C}
{Cromartie}, H.~T., {Fonseca}, E., {Ransom}, S.~M., {et~al.} 2020, NatAs, 4,
  72, \dodoi{10.1038/s41550-019-0880-2}

\bibitem[{{Dominik} {et~al.}(2012){Dominik}, {Belczynski}, {Fryer}, {Holz},
  {Berti}, {Bulik}, {Mandel}, \& {O'Shaughnessy}}]{2012ApJ...759...52D}
{Dominik}, M., {Belczynski}, K., {Fryer}, C., {et~al.} 2012, \apj, 759, 52,
  \dodoi{10.1088/0004-637X/759/1/52}

\bibitem[{{Essick} \& {Landry}(2020)}]{EssickLandry2020}
{Essick}, R., \& {Landry}, P. 2020, ApJ, 904, 80,
  \dodoi{10.3847/1538-4357/abbd3b}

\bibitem[{{Farr} \& {Chatziioannou}(2020)}]{FarrChatziioannou2020}
{Farr}, W.~M., \& {Chatziioannou}, K. 2020, RNAAS, 4, 65,
  \dodoi{10.3847/2515-5172/ab9088}

\bibitem[{{Farrow} {et~al.}(2019){Farrow}, {Zhu}, \& {Thrane}}]{FarrowZhu2019}
{Farrow}, N., {Zhu}, X.-J., \& {Thrane}, E. 2019, \apj, 876, 18,
  \dodoi{10.3847/1538-4357/ab12e3}

\bibitem[{{Fishbach} \& {Holz}(2020)}]{FishbachHolz2020}
{Fishbach}, M., \& {Holz}, D.~E. 2020, ApJL, 891, L27,
  \dodoi{10.3847/2041-8213/ab7247}

\bibitem[{{Fishbach} {et~al.}(2018){Fishbach}, {Holz}, \&
  {Farr}}]{2018ApJ...863L..41F}
{Fishbach}, M., {Holz}, D.~E., \& {Farr}, W.~M. 2018, \apjl, 863, L41,
  \dodoi{10.3847/2041-8213/aad800}

\bibitem[{{Foreman-Mackey} {et~al.}(2013){Foreman-Mackey}, {Hogg}, {Lang}, \&
  {Goodman}}]{ForemanMackeyHogg2013}
{Foreman-Mackey}, D., {Hogg}, D.~W., {Lang}, D., \& {Goodman}, J. 2013, PASP,
  125, 306, \dodoi{10.1086/670067}

\bibitem[{{Foreman-Mackey} {et~al.}(2014){Foreman-Mackey}, {Hogg}, \&
  {Morton}}]{ForemanMackeyHogg2014}
{Foreman-Mackey}, D., {Hogg}, D.~W., \& {Morton}, T.~D. 2014, ApJ, 795, 64,
  \dodoi{10.1088/0004-637X/795/1/64}

\bibitem[{{Fragos} {et~al.}(2019){Fragos}, {Andrews}, {Ramirez-Ruiz}, {Meynet},
  {Kalogera}, {Taam}, \& {Zezas}}]{2019ApJ...883L..45F}
{Fragos}, T., {Andrews}, J.~J., {Ramirez-Ruiz}, E., {et~al.} 2019, \apjl, 883,
  L45, \dodoi{10.3847/2041-8213/ab40d1}

\bibitem[{{Fryer} {et~al.}(2012){Fryer}, {Belczynski}, {Wiktorowicz},
  {Dominik}, {Kalogera}, \& {Holz}}]{2012ApJ...749...91F}
{Fryer}, C.~L., {Belczynski}, K., {Wiktorowicz}, G., {et~al.} 2012, \apj, 749,
  91, \dodoi{10.1088/0004-637X/749/1/91}

\bibitem[{{Galaudage} {et~al.}(2021){Galaudage}, {Adamcewicz}, {Zhu},
  {Stevenson}, \& {Thrane}}]{GalaudageAdamcewicz2021}
{Galaudage}, S., {Adamcewicz}, C., {Zhu}, X.-J., {Stevenson}, S., \& {Thrane},
  E. 2021, ApJL, 909, L19, \dodoi{10.3847/2041-8213/abe7f6}

\bibitem[{{Hogg} {et~al.}(2010){Hogg}, {Myers}, \& {Bovy}}]{HoggMyers2010}
{Hogg}, D.~W., {Myers}, A.~D., \& {Bovy}, J. 2010, ApJ, 725, 2166,
  \dodoi{10.1088/0004-637X/725/2/2166}

\bibitem[{{Kalogera} \& {Baym}(1996)}]{KalogeraBaym1996}
{Kalogera}, V., \& {Baym}, G. 1996, ApJL, 470, L61, \dodoi{10.1086/310296}

\bibitem[{{Kiziltan} {et~al.}(2013){Kiziltan}, {Kottas}, {De Yoreo}, \&
  {Thorsett}}]{KiziltanKottas2013}
{Kiziltan}, B., {Kottas}, A., {De Yoreo}, M., \& {Thorsett}, S.~E. 2013, \apj,
  778, 66, \dodoi{10.1088/0004-637X/778/1/66}

\bibitem[{{Kovetz} {et~al.}(2017){Kovetz}, {Cholis}, {Breysse}, \&
  {Kamionkowski}}]{KovetzCholis2017}
{Kovetz}, E.~D., {Cholis}, I., {Breysse}, P.~C., \& {Kamionkowski}, M. 2017,
  PhRvD, 95, 103010, \dodoi{10.1103/PhysRevD.95.103010}

\bibitem[{{Krishak} \& {Desai}(2020)}]{KrishakDesai2020}
{Krishak}, A., \& {Desai}, S. 2020, JCAP, 2020, 006,
  \dodoi{10.1088/1475-7516/2020/07/006}

\bibitem[{{Kruckow}(2020)}]{Kruckow2020}
{Kruckow}, M.~U. 2020, A\&A, 639, A123, \dodoi{10.1051/0004-6361/202037519}

\bibitem[{{Landry} {et~al.}(2020){Landry}, {Essick}, \&
  {Chatziioannou}}]{LandryEssick2020}
{Landry}, P., {Essick}, R., \& {Chatziioannou}, K. 2020, \prd, 101, 123007,
  \dodoi{10.1103/PhysRevD.101.123007}

\bibitem[{Liddle(2009)}]{Liddle2009}
Liddle, A.~R. 2009, ARNPS, 59, 95, \dodoi{10.1146/annurev.nucl.010909.083706}

\bibitem[{{LIGO Scientific Collaboration} {et~al.}(2015){LIGO Scientific
  Collaboration}, {Aasi}, {Abbott}, {Abbott}, {Abbott}, {Abernathy}, {Ackley},
  {Adams}, {Adams}, {Addesso}, \& et~al.}]{aLIGO}
{LIGO Scientific Collaboration}, {Aasi}, J., {Abbott}, B.~P., {et~al.} 2015,
  CQG, 32, 074001, \dodoi{10.1088/0264-9381/32/7/074001}

\bibitem[{{Loredo}(2004)}]{Loredo2004}
{Loredo}, T.~J. 2004, in Bayesian Inference and Maximum Entropy Methods in
  Science and Engineering: 24th International Workshop, Vol. 735, 195--206,
  \dodoi{10.1063/1.1835214}

\bibitem[{{Mandel}(2010)}]{Mandel2010}
{Mandel}, I. 2010, PhRvD, 81, 084029, \dodoi{10.1103/PhysRevD.81.084029}

\bibitem[{{Mandel} {et~al.}(2019){Mandel}, {Farr}, \& {Gair}}]{MandelFarr2019}
{Mandel}, I., {Farr}, W.~M., \& {Gair}, J.~R. 2019, \mnras, 486, 1086,
  \dodoi{10.1093/mnras/stz896}

\bibitem[{{Mandel} {et~al.}(2021){Mandel}, {M{\"u}ller}, {Riley}, {de Mink},
  {Vigna-G{\'o}mez}, \& {Chattopadhyay}}]{2021MNRAS.500.1380M}
{Mandel}, I., {M{\"u}ller}, B., {Riley}, J., {et~al.} 2021, \mnras, 500, 1380,
  \dodoi{10.1093/mnras/staa3390}

\bibitem[{{Mapelli} {et~al.}(2019){Mapelli}, {Giacobbo}, {Santoliquido}, \&
  {Artale}}]{2019MNRAS.487....2M}
{Mapelli}, M., {Giacobbo}, N., {Santoliquido}, F., \& {Artale}, M.~C. 2019,
  \mnras, 487, 2, \dodoi{10.1093/mnras/stz1150}

\bibitem[{{Most} {et~al.}(2020){Most}, {Papenfort}, {Weih}, \&
  {Rezzolla}}]{MostPapenfort2020}
{Most}, E.~R., {Papenfort}, L.~J., {Weih}, L.~R., \& {Rezzolla}, L. 2020,
  MNRAS, 499, L82, \dodoi{10.1093/mnrasl/slaa168}

\bibitem[{{Mukherjee} {et~al.}(2021){Mukherjee}, {Caudill}, {Magee}, {Messick},
  {Privitera}, {Sachdev}, {Blackburn}, {Brady}, {Brockill}, {Cannon},
  {Chamberlin}, {Chatterjee}, {Creighton}, {Fong}, {Godwin}, {Hanna},
  {Kapadia}, {Lang}, {Li}, {Lo}, {Meacher}, {Pace}, {Sadeghian}, {Tsukada},
  {Wade}, {Wade}, {Weinstein}, \& {Xiao}}]{2021PhRvD.103h4047M}
{Mukherjee}, D., {Caudill}, S., {Magee}, R., {et~al.} 2021, \prd, 103, 084047,
  \dodoi{10.1103/PhysRevD.103.084047}

\bibitem[{{{\"O}zel} \& {Freire}(2016)}]{OzelFreire2016}
{{\"O}zel}, F., \& {Freire}, P. 2016, ARA\&A, 54, 401,
  \dodoi{10.1146/annurev-astro-081915-023322}

\bibitem[{{{\"O}zel} {et~al.}(2012){{\"O}zel}, {Psaltis}, {Narayan}, \& {Santos
  Villarreal}}]{OzelPsaltis2012}
{{\"O}zel}, F., {Psaltis}, D., {Narayan}, R., \& {Santos Villarreal}, A. 2012,
  ApJ, 757, 55, \dodoi{10.1088/0004-637X/757/1/55}

\bibitem[{{Rhoades} \& {Ruffini}(1974)}]{RhoadesRuffini1974}
{Rhoades}, C.~E., \& {Ruffini}, R. 1974, PhRvL, 32, 324,
  \dodoi{10.1103/PhysRevLett.32.324}

\bibitem[{{Roulet} {et~al.}(2020){Roulet}, {Venumadhav}, {Zackay}, {Dai}, \&
  {Zaldarriaga}}]{RouletVenumadhav2020}
{Roulet}, J., {Venumadhav}, T., {Zackay}, B., {Dai}, L., \& {Zaldarriaga}, M.
  2020, PhRvD, 102, 123022, \dodoi{10.1103/PhysRevD.102.123022}

\bibitem[{{Santoliquido} {et~al.}(2021){Santoliquido}, {Mapelli}, {Giacobbo},
  {Bouffanais}, \& {Artale}}]{2021MNRAS.502.4877S}
{Santoliquido}, F., {Mapelli}, M., {Giacobbo}, N., {Bouffanais}, Y., \&
  {Artale}, M.~C. 2021, \mnras, 502, 4877, \dodoi{10.1093/mnras/stab280}

\bibitem[{{Shao} {et~al.}(2020){Shao}, {Tang}, {Jiang}, \&
  {Fan}}]{ShaoTang2020}
{Shao}, D.-S., {Tang}, S.-P., {Jiang}, J.-L., \& {Fan}, Y.-Z. 2020, PhRvD, 102,
  063006, \dodoi{10.1103/PhysRevD.102.063006}

\bibitem[{{Shi} {et~al.}(2012){Shi}, {Huang}, \& {Lu}}]{ShiHuang2012}
{Shi}, K., {Huang}, Y.~F., \& {Lu}, T. 2012, MNRAS, 426, 2452,
  \dodoi{10.1111/j.1365-2966.2012.21784.x}

\bibitem[{{Tews} {et~al.}(2021){Tews}, {Pang}, {Dietrich}, {Coughlin},
  {Antier}, {Bulla}, {Heinzel}, \& {Issa}}]{TewsPang2021}
{Tews}, I., {Pang}, P. T.~H., {Dietrich}, T., {et~al.} 2021, ApJL, 908, L1,
  \dodoi{10.3847/2041-8213/abdaae}

\bibitem[{{Thompson} {et~al.}(2019){Thompson}, {Kochanek}, {Stanek}, {Badenes},
  {Post}, {Jayasinghe}, {Latham}, {Bieryla}, {Esquerdo}, {Berlind}, {Calkins},
  {Tayar}, {Lindegren}, {Johnson}, {Holoien}, {Auchettl}, \&
  {Covey}}]{2019Sci...366..637T}
{Thompson}, T.~A., {Kochanek}, C.~S., {Stanek}, K.~Z., {et~al.} 2019, Sci, 366,
  637, \dodoi{10.1126/science.aau4005}

\bibitem[{{Thorsett} \& {Chakrabarty}(1999)}]{ThorsettChakrabarty1999}
{Thorsett}, S.~E., \& {Chakrabarty}, D. 1999, ApJ, 512, 288,
  \dodoi{10.1086/306742}

\bibitem[{{Usman} {et~al.}(2016){Usman}, {Nitz}, {Harry}, {Biwer}, {Brown},
  {Cabero}, {Capano}, {Dal Canton}, {Dent}, {Fairhurst}, {Kehl}, {Keppel},
  {Krishnan}, {Lenon}, {Lundgren}, {Nielsen}, {Pekowsky}, {Pfeiffer},
  {Saulson}, {West}, \& {Willis}}]{2016CQGra..33u5004U}
{Usman}, S.~A., {Nitz}, A.~H., {Harry}, I.~W., {et~al.} 2016, CQG, 33, 215004,
  \dodoi{10.1088/0264-9381/33/21/215004}

\bibitem[{{Vallisneri} {et~al.}(2015){Vallisneri}, {Kanner}, {Williams},
  {Weinstein}, \& {Stephens}}]{VallisneriKanner2015}
{Vallisneri}, M., {Kanner}, J., {Williams}, R., {Weinstein}, A., \& {Stephens},
  B. 2015, in JPhCS, Vol. 610, 012021, \dodoi{10.1088/1742-6596/610/1/012021}

\bibitem[{{Woosley} {et~al.}(2020){Woosley}, {Sukhbold}, \&
  {Janka}}]{2020ApJ...896...56W}
{Woosley}, S.~E., {Sukhbold}, T., \& {Janka}, H.~T. 2020, \apj, 896, 56,
  \dodoi{10.3847/1538-4357/ab8cc1}

\end{thebibliography}
\bibliographystyle{aasjournal}
\end{document}